\documentclass{ifacconf}

\usepackage{graphicx}      
\usepackage{natbib} 
\usepackage{amsmath}
\usepackage{tikz}
\usepackage{graphicx}
\usepackage{amssymb} 
\usepackage[version=4]{mhchem}
\usepackage{siunitx}
\usepackage{longtable,tabularx}
\usepackage{float}
\usepackage{caption}
\usepackage{subcaption}
\usepackage{booktabs}
\usepackage{tikz-network}
\usetikzlibrary{arrows.meta}
\usepackage{algorithm}
\usepackage[noend]{algpseudocode}

\begin{document}
\begin{frontmatter}

\title{Resilient Multi-Agent Target Localization and Tracking via BAG-Aware Mutual Information Maximization under False Data Injection Attacks}

\thanks[footnoteinfo]{The authors gratefully acknowledge the support from ARO Grant W911NF-23-1-0174}

\author[First]{Bibek Adhikari} 
\author[First]{Samrat Chaulagain} 
\author[First]{Kamesh Subbarao}

\address[First]{Department of Mechanical and Aerospace Engineering,\\ The University of Texas at Arlington, Arlington, TX 76019 USA (e-mail: bibek.adhikari@mavs.uta.edu, sxc9808@mavs.uta.edu, subbarao@uta.edu).}

\begin{abstract}               
Cooperative multi-agent networks deployed for target localization and tracking remain critically vulnerable to malicious cyber attacks, since a single compromised agent can corrupt the centralized target belief and mislead the estimation process across the entire network. This paper presents a resilient multi-agent target localization and tracking framework under a limited agent sensing range and false-data injection (FDI) attacks. The proposed method combines information-based navigation for target search and posterior target state uncertainty reduction, Bayesian attack graph (BAG) based attack detection for probabilistic identification of the compromised agents, and reachable set-guided recovery that guides the uncompromised agents to relocalize the target when the sole localizing agent is attacked. The framework preserves the network's localization capability after an attack by trust-gating compromised sensing information and guiding attack-immune agents toward a dynamically constructed recovery set representing the new target belief using a Linear Quadratic Regulator (LQR). Simulation validates that the proposed framework maintains a robust tracking performance under FDI attacks, demonstrating the potential for resilient autonomy in safety-critical applications.

\end{abstract}


\end{frontmatter}

\section{Introduction}
Cooperative multi-agent network have advanced in autonomous systems in the field of localization and tracking. In recent years, such networks have been employed in applications such as autonomous surveillance \citep{drones}, environmental monitoring \citep{engproc}, and threat localization and neutralization \citep{saeed}. Target tracking through a multi-agent network enables the system to share information among agents via a centralized or decentralized framework. Localization under an agent's sensing range constraint introduces complexity to the network, requiring the target position to be inferred cooperatively from the combination of presence and absence of observations. Under these conditions, agents must actively coordinate their movement to gradually reduce a shared uncertainty of the target state, scanning the area until enough observations are obtained to localize the target with certainty. Particle filter-based approach for cooperative localization is a well-suited framework for systems with unknown target motion models and nonlinear measurements, relying solely on weighted information about the target state \citep{Hoffmann}. However, when these systems are deployed in adversarial environments, they become vulnerable to cyber-physical attacks such as sensor tampering, false data injection, and Denial-of-Service (DoS) attacks \citep{cpa}.\\
In particular, FDI attacks represent a threat where adversaries inject malicious data into the system, often by taking control of certain agents or by tampering with their communication channels, resulting in invalid measurements. The system with centralized belief is highly vulnerable to these attacks, especially since a single compromised agent can corrupt the entire distribution of the target state, which propagates to other agents in the network. In particle-filter estimation of the target state under FDI attacks, the attacked agents' target estimates could be treated as trusted positions, leading the entire network to pursue the ghost target. The situation is further complicated if the sole initial localizing agent is attacked by the malicious target.\\
To mitigate cyber-physical attacks, recent research has focused on secure estimation and resilient filtering strategies that aim to detect and isolate compromised nodes \citep{Chen}. Bayesian attack graphs provide a probabilistic framework for modeling the propagation of cyber attacks across a network by exploiting dependencies among nodes and have been used for security assessment and inference of compromise in networked systems  \citep{BAG1, BAG2}. Recent work has further developed BAG-based MMSE monitoring and attack detection under imperfect observations \citep{kazeminajafabadi2023optimal}, as well as BAG-aware resilient target tracking in multi-agent systems \citep{samratBAG}.\\ 
The rest of the paper is formulated as follows. Section 2 presents the objective and problem statement, while Section 3 presents modeling and analysis of the graph network, target, sensing agent, and measurement model. Section 4 formulates the methodology of the work. The simulation and experimental setups are described in Section 5, and the paper concludes in Section 6.

\section{Problem Statement}

The paper addresses a multi-agent network tasked with localizing and tracking a dynamic intruder target in a bounded search space. The agents are equipped with a sensor of limited measuring range. A particle filter is employed in the system to represent the target belief and estimate the target state based on the measurements from the sensing agents. The mutual information between the target and the sensor observation is used to navigate inside the search space to localize the target. The control objective of the system is to compute a control input that arranges agents’ positions, maximizing the mutual information, which leads to future observations that minimize the target uncertainty. \\ 
The paper further addresses a resilient cooperative localization and tracking of the target in the presence of false data injection into the localizing agent through the target. The challenge arises when the sole localizing agent gets compromised, providing biased or inconsistent observation, jeopardizing the system's particle belief. The mission objective is to localize the target under the influence of FDI while preventing the attack from degrading the system's performance, and continue to preserve a bounded posterior belief around the moving target, corresponding to persistent tracking. The contribution of the paper is a unified resilient framework that combines mutual information-based target localization, Bayesian attack graph-based compromise detection, and reachable-set-guided recovery for robust relocalization and tracking.

\section{System Modeling and Analysis}
\label{ch:sysmodel}
The global coordinate system is a fixed reference frame represented as $ I = \{ O_{I},X_I,Y_I, Z_I \} $.  

\subsection{Target Model}
Let $\mathbf{x}_T = \begin{bmatrix} x_T & y_T & \dot{x}_T & \dot{y}_T\end{bmatrix}^\top$ be the state of the target in inertial frame in the search domain $\mathcal{X}$, such that $\mathbf{x}_T$ $\in$ $\mathcal{X}$, whose model is represented as:
\begin{equation}
    \dot{\mathbf{x}}_T(t) = \mathbf{A}_T \: \mathbf{x}_T(t) + \mathbf{B}_T \:  \boldsymbol{\nu}_T(t), 
\end{equation}
with
\[
\mathbf{A}_T =\begin{bmatrix}
0 & 0 & 1 & 0\\
0 & 0 & 0 & 1\\
0 & 0 & 0 & 0\\
0 & 0 & 0 & 0
\end{bmatrix} \in \mathbb{R}^{4\times 4},
\qquad
\mathbf{B}_T =\begin{bmatrix}
0 & 0\\
0 & 0\\
1 & 0\\
0 & 1
\end{bmatrix} \in \mathbb{R}^{4\times 2},
\]
where $\mathbf{A}_T$ and $\mathbf{B}_T$ are the system and input matrix of the target model, respectively, at time $t$, and $\boldsymbol{\nu}_T$ is the random acceleration disturbance such that $\boldsymbol{\nu}_T \sim \mathcal{N}\bigl(\mathbf{0},\boldsymbol\Sigma_{\boldsymbol{\nu}_T} \bigr) \in \mathbb{R}^{2}$ is modeled as the zero-mean Gaussian noise with covariance $\boldsymbol\Sigma_{\boldsymbol{\nu}_T}$. The position of the target is denoted in the inertial frame as $\mathbf{p}_T = \begin{bmatrix} x_T & y_T\end{bmatrix}^\top$. A prior distribution of the target $p(\mathbf{x}_T(0))$ is known, around which the search domain is created. 

\subsection{Sensing Agent Model}
Consider a set of $A$ sensing agents localizing the target and let $\mathbf{x}_{i} = \begin{bmatrix} x_{i} & y_{i} & \dot{x}_{i} & \dot{y}_{i} \end{bmatrix}^\top \in \mathbb{R}^4$ be the state of $i^\mathrm{th}$ agent in inertial frame with dynamics given as:
\begin{equation}
    \label{agent dynamics}
    \dot{\mathbf{x}}_{i}(t) = \mathbf{A} \mathbf{x}_{i}(t)  + \mathbf{B} \mathbf{u}_{i}(t),
\end{equation}
where $\mathbf{A}$ and $\mathbf{B}$ are the system and input matrices of sensing agents respectively, and $\mathbf{u}_{i}(t)$ is the control input of $i^\mathrm{th}$ agent at time $t$ with
\begin{equation*}
    \mathbf{A} = \begin{bmatrix} 0 & 0 & 1 & 0 \\ 0 & 0 & 0 & 1 \\ 0 & 0 & 0 & 0 \\ 0 & 0 & 0 & 0\end{bmatrix},\mathbf{B} = \begin{bmatrix} 0 & 0 \\ 0 & 0 \\ 1 & 0 \\ 0 & 1 \end{bmatrix}, \mathbf{u}_i(t) = \begin{bmatrix}a_{x_i} \\ a_{y_i} \end{bmatrix},
\end{equation*}
where $a_{x_i}$  and $a_{y_i}$ are the acceleration control inputs. The position of the $i^\mathrm{th}$ agent is denoted in the inertial frame as $\mathbf{p}_i = \begin{bmatrix} x_i & y_i\end{bmatrix}^\top$, whereas its heading is denoted as $\psi_i$ about the $Z_I$ axis.

\subsection{Measurement Model}
Let $\mathbf{z}_i$ be the measurement made by the $i^\mathrm{th}$ sensing agent, measuring the range from the agent to the target, and its bearing. The measurements of the target are obtained only when the target lies within the sensor’s maximum range. The range measurement is represented as: 
\begin{align}
\label{eq:meas_r}
    z^r_i(k) &= \|\mathbf{p}_i(k) - \mathbf{p}_T(k) \| + \epsilon^r_i (k), \\
    \epsilon^r &\sim \mathcal{N}(0,\sigma^2_r).   
\end{align}
The bearing measurement from the agent to the target is expressed as:
\begin{align}
\label{eq:meas_theta}
    z^\theta_i(k) &= \arctan \bigg( \frac{y_T(k)-y_{i}(k)}{x_T(k)-x_{i}(k)} \bigg) + \epsilon^\theta_i (k), \\
    \epsilon^\theta &\sim \mathcal{N}(0,\sigma^2_\theta),    
\end{align}
where $\epsilon^r_i (k) $ and $\epsilon^\theta_i (k) $ are the range and bearing measurement noise of $i^\mathrm{th}$ sensing agent respectively at time step $k$, with the Gaussian distribution of zero-mean and variance $\sigma^2_r$ and $\sigma^2_\theta$ respectively. The aggregate measurement for $i^\mathrm{th}$ sensing agent at time step $k$ is represented as $\mathbf{z}_i(k)= \begin{bmatrix} z^r_i(k) & z^\theta_i(k) \end{bmatrix}^\top \in \mathbb{R}^2$.
The particle-wise likelihood used for belief update incorporates both range and bearing measurements, which is the probability of the current measurement conditioned on the $n^{\mathrm{th}}$ particle being the true target state. Assuming independent measurement noise, the likelihood of the $n^\mathrm{th}$ particle for the $i^\mathrm{th}$ agent is expressed as:
\begin{equation}
    L^n_i(k) = p(\mathbf{z}_i(k) | \mathbf{x}_T = \tilde{\mathbf{x}}_{T}^n(k)),
\end{equation}
where $\tilde{\mathbf{x}}_{T}^n$ denote the state of the particle.

\subsection{False Attack Injection Model}
The attack is triggered when an agent breaches the target's attack range ($R_{\text{atk}}$). It is assumed that the target's attack range is smaller than the agents' sensing range. Once an agent is compromised, false data is injected into its sensor, which is additive to its true measurement. The measurement of the compromised agent is modeled as:
\begin{equation}
    \label{compromised measurement}
    z^r_{i,\mathrm{atk}}(k)  = z^r_i(k) + \alpha_i(k), 
\end{equation}
where $\alpha_i(k)\sim \mathcal{N}(b,\sigma^2_\alpha)$ denotes the false-data injected into compromised agent, with mean $b$ and variance $\sigma^2_\alpha$.

\section{Methodology}

\subsection{Mutual Information Maximization}

Mutual information is evaluated between the target state and range measurement. The measurement entropy is approximated using a discretized grid $z^r \in [z^r_\text{min} ,z^r_\text{max}]$ with spacing $\Delta z^r$. The optimal control input to plan the next agent state is calculated by optimizing mutual information represented as:
\begin{equation}
\mathbf{u}_{m_i}^{\star}(k)
=
\arg\max_{\mathbf{u}_{m_i}\in\mathcal{U}_i}
I\!\left(\mathbf{x}_T;z^r_i(k)\mid \mathcal{Z}^r_{k-1} \right),
\label{control MIM}
\end{equation}
with
\begin{equation}
\label{eq:Iform}
I(\mathbf{x}_T; z^r_{i}(k) \mid \mathcal{Z}^r_{k-1} ) = H(z^r_{i}(k) \mid \mathcal{Z}^r_{k-1}) - H(z^r_{i}(k) \mid \mathbf{x}_T, \mathcal{Z}^r_{k-1}), 
\end{equation}
where $\mathcal{Z}^r_{k-1} = \{\mathbf{z}^r(0),...,\mathbf{z}^r(k-1)\}$ represents the range measurement history up until $k$. The entropy of measurement given the measurement history and entropy of measurement given the target state and measurement history are respectively expressed as:
\begin{equation}
\label{eq:Hz}
\begin{aligned}
    H(z^r_{i}(k) \mid \mathcal{Z}^r_{k-1}) &= - \sum_{z^r_i \in \mathbf{\mathcal{Z}^r}} \bigg( \bigg( \sum_{n=1}^N  w^n (k) \, l^n_i(k) \bigg) . \\ 
    & \cdot \log \bigg( \sum_{n=1}^N  w^n (k) \, l^n_i(k)   \bigg) \bigg) \Delta z^r,
\end{aligned}
\end{equation}
\begin{equation}
\label{eq:HZgX4}
\begin{aligned}
    H(z^r_{i}(k) \mid \mathbf{x}_T(k), \, \mathcal{Z}^r_{k-1})  &= -\sum_{z^r_i \in \mathbf{\mathcal{Z}^r}} \bigg( \sum_{n=1}^N \bigl( w^n (k) \, l_i^n(k). \\ 
    & \cdot \log \; \left( l_i^n(k)  \right) \bigr) \bigg) \, \Delta z^r,
\end{aligned}
\end{equation}
where $l^n_i = p(z^r_i | \mathbf{x}_T = \tilde{\mathbf{x}}_{T}^n)$ is the particle-wise measurement likelihood used for control step utilizing range measurement, and $w^n$ is the weight of the $n^{th}$ particle. The derivation of the respective entropy terms are presented in our earlier work \citep{BibekMIM}.

\subsection{Localization of target}

The differential Gaussian entropy of the target state conditioned on the measurement history is used to quantify the uncertainty of the target probability distribution, which is expressed as:
\begin{equation}
\label{eq:H}
H(\mathbf{x_T} \mid \mathcal{Z}^r_k) = \frac{1}{2}\log\!\Big((2\pi e)^{d}\det(\boldsymbol{\Sigma}_k)\Big),
\end{equation}
where $e$ and $d$ denote the Euler's number and dimension of the target state respectively, and $\boldsymbol{\Sigma}$ represents the posterior covariance estimated from the particle belief. \\
The posterior belief of the target state increasingly concentrates as the agents receive informative observations, leading to a decrease in posterior differential Gaussian entropy. Once the entropy drops below a predetermined threshold corresponding to an acceptable degree of uncertainty, the target can be deemed localized, and converging to a constant value below the threshold signifies the active tracking of the target. Subsequent measurements are then used to maintain a bounded posterior belief around the target state, corresponding to persistent tracking.\\

\subsection{Bayesian Attack Graph Network}
To capture cyber-attack propagation over the multi-agent network, Bayesian attack graph is employed. The BAG is represented by a tuple, represented as: $$\mathcal{G}_{\text{BAG}} = (\mathcal{V,E,T,P}),$$
where, $\mathcal{V} = \{1,2,\dots,A\}$ is the set of sensing agents, $\mathcal{E}\subseteq\mathcal{V}\times\mathcal{V}$ is the set of directed edges, $\mathcal{T}$ denotes the node types, and $\mathcal{P}$ denotes the exploit probabilities. An edge $(i,j)\in\mathcal{E}$ indicates that compromise of agent $i$ can lead to compromise of agent $j$. Each agent has binary compromise state $s_j(k)\in\{0,1\}$, and the network state is
\begin{equation}
    \mathbf{s}(k) = \begin{bmatrix}
        s_1(k) & s_2(k) & \cdots & s_A(k)
    \end{bmatrix}^\top \in \{0,1\}^A
    \label{BAG-compromise state at time k}.
\end{equation}
A subset $\mathcal{V}_c \subseteq \mathcal{V}$ is vulnerable to direct external attack, where each $j \in \mathcal{V}_c$ is assigned external exploit probability $\rho_j$. Each edge $(i,j)\in\mathcal{E}$ is assigned internal exploit probability $\rho_{ij}$. Nodes are classified as AND or OR types. The BAG evolves as the binary-state Markov process, and the conditional probability of node $j$ depends on its node type and the previous compromise state $\mathbf{s}(k-1)$. Once compromised, a node is assumed to remain compromised for the rest of the mission. To infer the hidden compromise states, binary monitoring observations are introduced. Let $y_j(k) \in \{0,1\}$ denote the monitoring output for agent $j$, where $y_j(k) = 1$ indicates a detected compromise. The exploit probability calculation and observation model follows \citep{samratBAG}.\\
Compromised agents may be missed with probability $q_{\text{miss}}$, while healthy agents do not generate false alarms. The joint observation vector is
$$\mathbf{y}(k) = \begin{bmatrix}
    y_1(k) & y_2(k) &\cdots & y_A(k)
\end{bmatrix}^\top.$$
Let \begin{equation}
    \mathbf\Omega = \begin{bmatrix}
        \mathbf{s}^1,\mathbf{s}^2,\dots,\mathbf{s}^{2^A}
    \end{bmatrix} \in \{0,1\}^{A\times 2^A}
    \label{BAG all states}
\end{equation}
denote the set of all possible network compromise configurations. The posterior and predictive belief vectors are defined as:
\begin{equation}
\begin{aligned}
    \mathbf\Pi_{k|k}(i) &= p\bigl( \mathbf{s}(k) = \mathbf{s}^i \mid \mathbf{y}(1:k)\bigr), \\
    \mathbf\Pi_{k|k-1}(i) &= p\bigl( \mathbf{s}(k) = \mathbf{s}^i \mid \mathbf{y}(1:k-1)\bigr), \\
\end{aligned}
\label{BAG update predict}
\end{equation}
for $i = 1,2,\dots,2^A$. The state transition matrix $\mathbf{M}_k \in \mathbb{R}^{2^A \times 2^A}$ is
\begin{equation}
    \label{BAG state transition matrix}
    (\mathbf{M}_k)_{ij} = p\bigl( \mathbf{s}(k) = s^i \mid \mathbf{s}(k-1) = s^j \bigr),
\end{equation}
which gives the predictive update
\begin{equation}
    \label{BAG-predictive}
    \mathbf\Pi_{k|k-1} = \mathbf{M}_k \mathbf\Pi_{k-1|k-1}.
\end{equation}
Given $\mathbf{y}(k)$ at time $k$, the likelihood vector is
\begin{equation}
    \label{BAG-likelihood vector}
    \bigl(\mathbf{T}_k(\mathbf{y}(k))\bigr)_i = p \bigl( \mathbf{y}(k) \mid \mathbf{s}(k) = \mathbf{s}^i \bigr),
\end{equation}
and the posterior update is
\begin{equation}
    \label{BAG-posterior update}
    \mathbf\Pi_{k|k} = \dfrac{\mathbf{T}_k \bigl(\mathbf{y}(k)\bigr) \circ \mathbf\Pi_{k|k-1}}{\| \mathbf{T}_k \bigl(\mathbf{y}(k)\bigr) \circ \mathbf\Pi_{k|k-1}\|_1},
\end{equation}
where $\circ$ denotes the Hadamard (element-wise) product and $\|\cdot\|_1$ denotes the $\ell_1$-norm.
The optimal minimum mean-square-error estimate of the compromise state is
\begin{equation}
    \label{BAG-compromise detection}
    \hat{\mathbf{s}}^{\text{MMSE}}(k) = \text{round}\bigl( \mathbf\Omega \mathbf\Pi_{k|k} \bigr),
\end{equation}
where the round$(\cdot)$ maps each entry to the nearest binary value~\citep{kazeminajafabadi2023optimal}. This estimate defines the trust variable used in the resilient particle-filter update as:
\begin{equation}
    \label{BAG-trust coefficient}
    \gamma_i(k) = \begin{cases}
        1, & \hat{s}_i^{\text{MMSE}}(k) = 0,\\
        0, & \hat{s}_i^{\text{MMSE}}(k) = 1.
    \end{cases}
\end{equation}

\subsection{Target Reachable Region Approximation}
Once the BAG detects an attack at time $k_d$ a recovery set is initialized using the last trusted target-belief mean prior to the detected compromise. The last trusted instant is estimated retrospectively from the stored range-residual history of the compromised agent as:
\begin{equation}
\hspace*{-0.2cm}r_i(k)=\left|z^r_i(k)-\bar{z}^r_i(k)\right|, \quad
k_t=\max\{k<k_d : r_i(k)<r_{\mathrm{th}}\},
\label{trusted iter}
\end{equation}
where $r_i(k)$ denotes the residual of $i^\mathrm{th}$ agent, $r_{\mathrm{th}}$ is the residual threshold, and $\bar{z}^r$ is the estimated range measurment from the particle mean. Assuming the target speed is bounded by $v_T^\mathrm{max}$, the feasible post-attack target reachable position is represented by a reachable set as:
\begin{equation}
\label{recovery set}
\mathcal{R}(k) = \Big\{ \mathbf{p}_T \in \mathbb{R}^2 : \| \mathbf{p}_T - \boldsymbol{\mu}(k_t) \| \leq R_0 + \big( (k-k_t)v_T^{\max}\Delta t \big)\Big\},
\end{equation}
where $\boldsymbol{\mu}(k_t)$ denotes the target-belief mean, and $R_0$ is the initial recovery radius reflecting uncertainty at the last trusted instant. A new set of particles is generated inside a set representing the feasible target position post attack detection.

\subsection{Linear Quadratic Regulator Guidance} 
A LQR guidance law is used when an agent is outside an informative sensing configuration, particularly during particle-sparse event and recovery-set entry phase. During particle-sparse event, agents' lookahead state is the particle-belief mean. For recovery-set entry phase, recovery agents are guided into the set with particle-belief mean being the lookahead point. Once sufficient local particle support is observed, the agents switch from LQR guidance to mutual-information-maximizing search within the recovery region. Similarly, during this phase, the desired state for compromised agents is selected to move the agent away from the target attack region. The guidance law is represented as:
\begin{equation}
\mathbf{u}_{g_i}(k) = -\mathbf{K}_g \big( \mathbf{x}_i(k)-\mathbf{x}_{d,i}(k) \big),
\label{control lqr}
\end{equation}
where $\mathbf{x}_{d_i} = \begin{bmatrix} x_{d_i} & y_{d_i} & \dot{x}_{d_i} & \dot{y}_{d_i}\end{bmatrix}^\top$ is the desired state composed of lookahead position $\mathbf{p}_{d_i}$ and zero desired velocity, and $\mathbf{K}_g$ is LQR feedback gain obtained from discrete-time LQR design. The inter-agent and agent-target collision is avoided using the collision avoidance constraint \citep{BibekMIM}.

\subsection{Trust-Gated Particle Filter Update}
Following attack detection, the particle filter update is conditioned on the agent’s trust status. Measurements from trusted agents are used to evaluate the likelihood, while those from compromised agents are excluded. As a result, posterior weight updates and particle resampling rely only on trusted agents, thereby preventing the corrupted data from degrading the target belief and enabling relocalization and tracking of the target using trusted agents. The normalized particle weights are updated as:
\begin{equation}
\label{trusted particle update}
w^n(k) = w^n(k-1) \displaystyle\prod_{i=1}^{A} (L_i^n(k))^{\gamma_i(k)}.
\end{equation}

\begin{algorithm}[t]
\caption{Multi-Agent Target Localizaton}
\label{alg:bag_resilient_localization}
\begin{algorithmic}[1]
\State \textbf{Inputs:} Search Space $\mathcal{X}$, $\mathcal{G}_{\text{BAG}}$, initial BAG belief $\Pi_{1|0}$, initial agent states $\mathbf{x}_i(0)$, entropy threshold $H_{\mathrm{th}}$.
\For{$k=1,2,\dots$}
    \If {$z_i^r(k) \leq R_\text{atk}$} 
        \State initiate attack
    \EndIf
    \State Propagate attack using BAG
    \State Obtain sensor measurements: $\mathbf{z}_i(k)$
    \State Observe cyber monitoring output $\mathbf{y}(k)$ 
    \State BAG update: $\Pi_{k|k}$ using Eq.~(\ref{BAG-posterior update})
    \State BAG prediction: $\Pi_{k+1|k}$ using Eq.~(\ref{BAG-predictive})
    \State BAG attack detection: $\hat{s}^{\mathrm{MMSE}}(k)$ using Eq.~(\ref{BAG-compromise detection})
    \State Assign trust: $\gamma_i(k)$ using Eq.~(\ref{BAG-trust coefficient})
    \State Update particle weights: $w_n(k)$ using Eq.~(\ref{trusted particle update})
    \State Compute posterior entropy: $H(\mathbf{x}_T|\mathcal{Z}^r_k)$ using Eq.~(\ref{eq:H})

    \If{$H(\mathbf{x}_T|\mathcal{Z}^r_k)<H_{\mathrm{th}}$ for $\tau$ consecutive steps}
        \State Declare target localized
        \State Continue posterior update for target tracking
    \EndIf
    \If{attack detected}
        \State Estimate $k_t$ using Eq.~(\ref{trusted iter})
        \State Construct $\mathcal{R}(k)$ using Eq.~(\ref{recovery set})
        \If{ $i$ is a compromised agent}
           \State Control update: $\mathbf{u}_{g_i}(k)$ using Eq.~(\ref{control lqr}) with $\mathbf{p}_{d,i}(k) \gets \mathbf{p}_i(0)$
        \EndIf
    \EndIf
    \State Agent control update:
            \If{sufficient local particle available} $\mathbf{u}_{m_i}(k)$ using Eq.~(\ref{control MIM}) 
            \Else \:$\mathbf{u}_{g_i}(k)$ using Eq.~(\ref{control lqr}) with
                \If {attack detected}\:$\mathbf{p}_{d,i}(k) \gets \boldsymbol\mu(k_t)$ 
                \Else\:$\mathbf{p}_{d,i}(k) \gets \boldsymbol\mu(k)$
                \EndIf
            \EndIf
\EndFor
\end{algorithmic}
\end{algorithm}

\section{Simulation and Results}

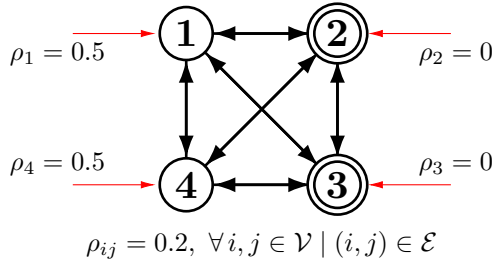
\begin{figure}[ht]
    \centering
    \begin{tikzpicture}
       \SetVertexStyle[MinSize=0.7\DefaultUnit,TextFont={\bfseries\normalsize},FillColor = white, TextColor=black]
        \SetEdgeStyle[LineWidth=1pt, Color=black,Arrow=-{Latex[length=3mm,width=2mm]}]
        \Vertex [x=0, y=0,IdAsLabel,fontscale=1.5]{4}
        \Vertex [style={double,double distance=1.5pt},x=2, y=2,IdAsLabel,fontscale=1.5]{2}
        \Vertex [x=0, y=2,IdAsLabel,fontscale=1.5]{1}
        \Vertex [style={double,double distance=1.5pt},x=2, y=0,IdAsLabel,fontscale=1.5]{3}
        
        \Edge[Direct](1)(2)
        \Edge[Direct](1)(3)
        \Edge[Direct](1)(4)
        \Edge[Direct](2)(1)
        \Edge[Direct](2)(3)
        \Edge[Direct](2)(4)
        \Edge[Direct](3)(1)
        \Edge[Direct](3)(2)
        \Edge[Direct](3)(4)
        \Edge[Direct](4)(2)
        \Edge[Direct](4)(3)
        \Edge[Direct](4)(1)

        \draw[red,-{Latex[length=2mm,width=1mm]}]
            (-1.5,2) -- (-0.4,2)
            node[midway, below left,text=black] {$\rho_1 = 0.5$};

        \draw[red,-{Latex[length=2mm,width=1mm]}]
            (3.5,2) -- (2.4,2)
            node[midway, below right,text=black] {$\rho_2 = 0$};

        \draw[red,-{Latex[length=2mm,width=1mm]}]
            (3.5,0) -- (2.4, 0)
            node[midway, above right,text=black] {$\rho_3 = 0$};

        \draw[red,-{Latex[length=2mm,width=1mm]}]
            (-1.5,0) -- (-0.4,0)
            node[midway, above left,text=black] {$\rho_4 = 0.5$};
        \node at (1, -0.8) {$\rho_{ij}=0.2,\ \forall\, i,j\in\mathcal{V} \mid (i,j)\in\mathcal{E}$};
    \end{tikzpicture}
    \caption{Communication network of the four-agent system. Double-circled agents denote AND nodes, while single-circled agents denote OR nodes.}
    \label{fig:communication network} 
\end{figure}

\begin{figure}[t]
    \centering
    \begin{subfigure}[b]{0.18\textwidth}
        \centering
         \includegraphics[trim={0cm 0.05cm 0.15cm 0.14cm},clip,width=1\linewidth]{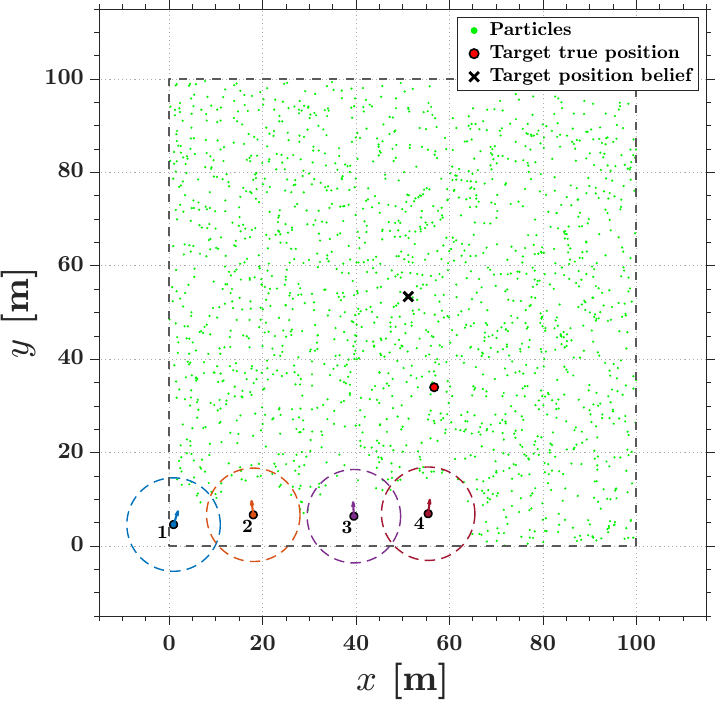}
        \caption{t = 5 secs}
        \label{fig:s1}
    \end{subfigure}
    \hspace{0.015\textwidth}
    \begin{subfigure}[b]{0.18\textwidth}
        \centering
        \includegraphics[trim={0cm 0.05cm 0.15cm 0.14cm},clip,width=1\linewidth]{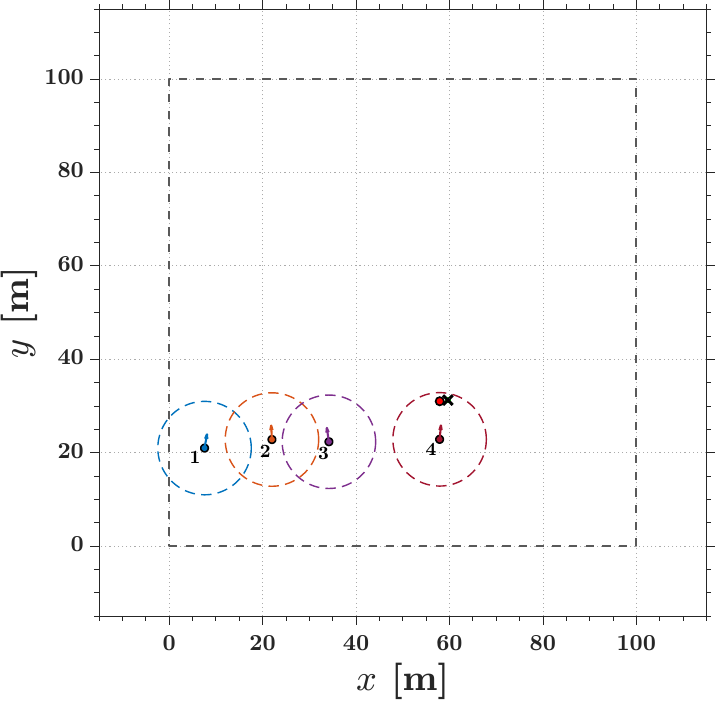}
        \caption{t = 14 secs}
        \label{fig:s2}
    \end{subfigure}
    
    \begin{subfigure}[b]{0.18\textwidth}
        \centering
        \includegraphics[trim={0cm 0.05cm 0.15cm 0.14cm},clip,width=1\linewidth]{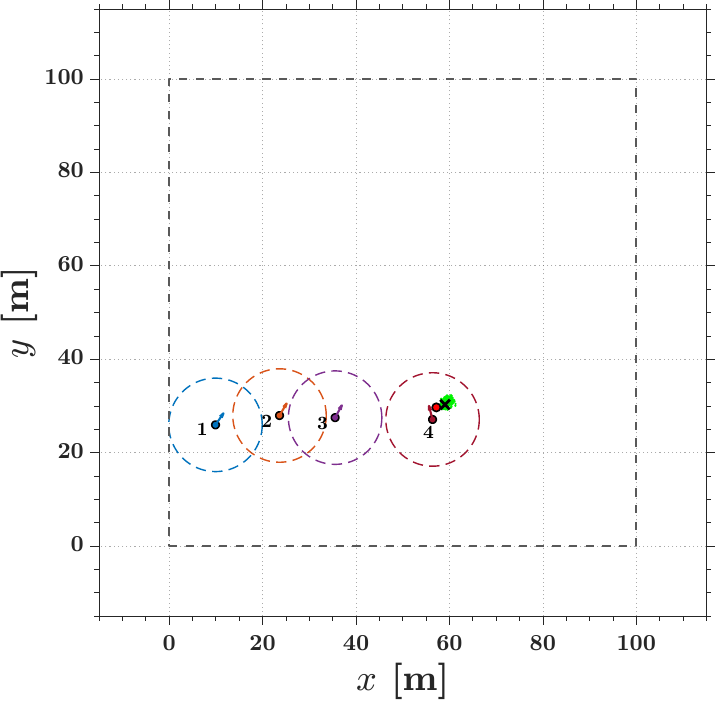}
        \caption{t = 17 secs}
        \label{fig:s3}
    \end{subfigure}
    \hspace{0.015\textwidth}
    \begin{subfigure}[b]{0.18\textwidth}
        \centering
        \includegraphics[trim={0cm 0.05cm 0.15cm 0.14cm},clip,width=1\linewidth]{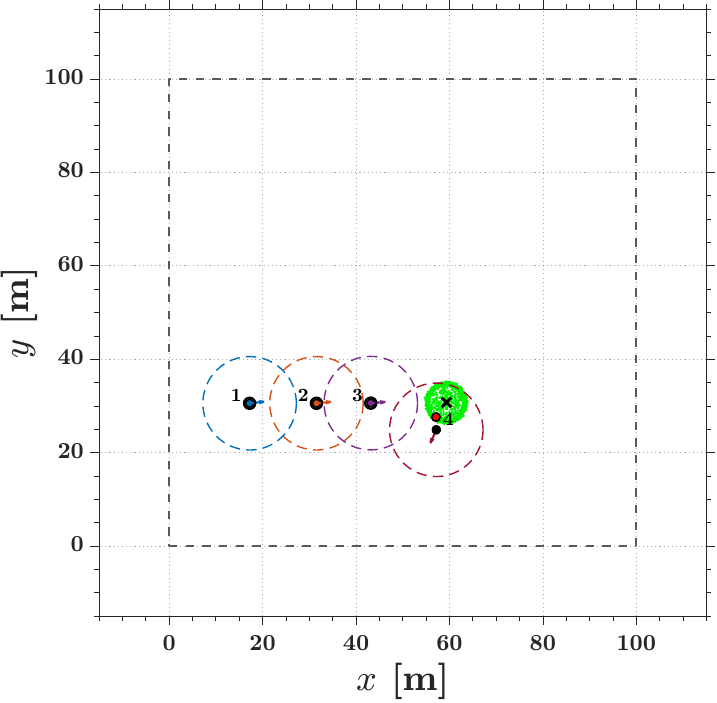}
        \caption{t = 22 secs}
        \label{fig:s4}
    \end{subfigure}
    
    \begin{subfigure}[b]{0.18\textwidth}
        \centering
        \includegraphics[trim={0cm 0.05cm 0.15cm 0.14cm},clip,width=1\linewidth]{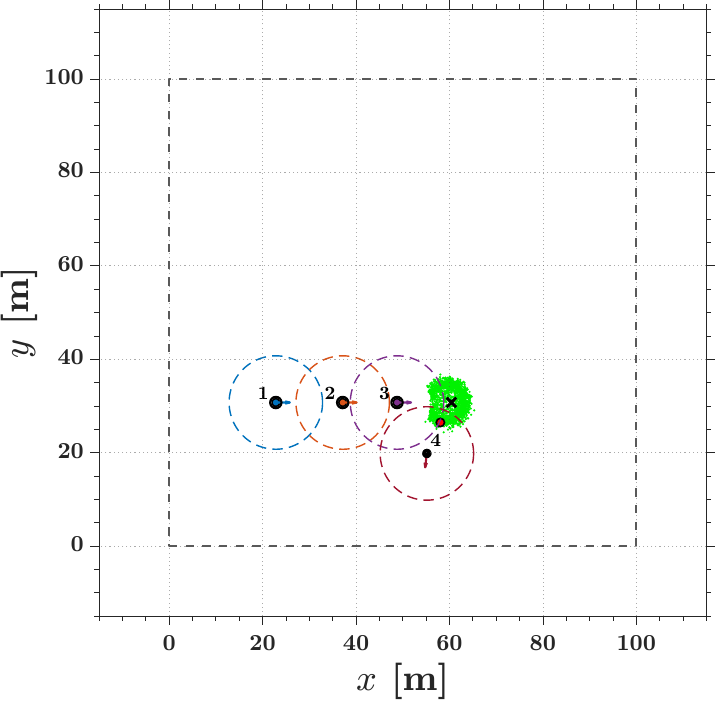}
        \caption{t = 25 secs}
        \label{fig:s5}
    \end{subfigure}
    \hspace{0.015\textwidth}
    \begin{subfigure}[b]{0.18\textwidth}
        \centering
        \includegraphics[trim={0cm 0.05cm 0.15cm 0.14cm},clip,width=1\linewidth]{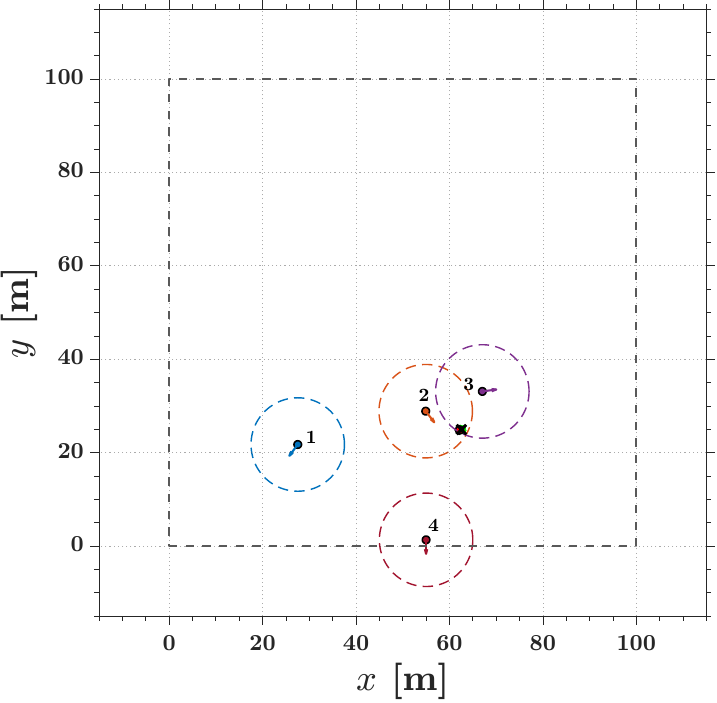}
        \caption{t = 35 secs}
        \label{fig:s6}
    \end{subfigure}

\caption{Mission evolution for target localization using MAN under FDI-attacks: (a) Target search, (b) Initial target localization, (c) FDI attack onset, (d) Recovery-set generation, (e) Post-set recovery maneuver, and (f) Cooperative relocalization. The colored circles represent the agents, and surrounding dashed circles indicate their maximum sensing ranges.}
\label{fig:snaps}
\end{figure}

\begin{figure*}[!t]
\centering
\begin{minipage}[t]{0.32\textwidth}
    \centering
    \includegraphics[width=\linewidth]{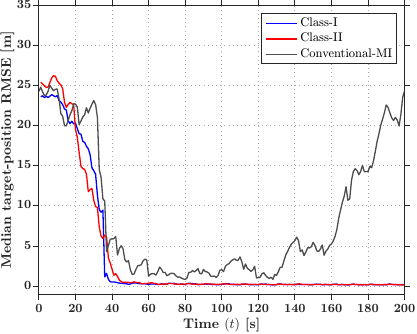}
    \caption{Median target-position RMSE}
    \label{fig:RMSE}
\end{minipage}\hfill
\begin{minipage}[t]{0.32\textwidth}
    \centering
    \includegraphics[width=\linewidth]{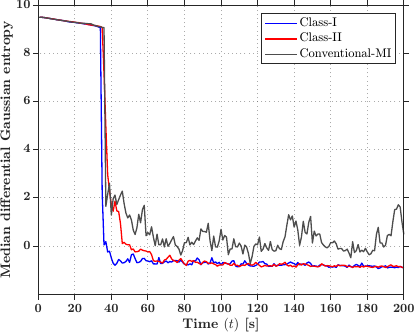}
    \caption{Median posterior differential Gaussian entropy of target belief}
    \label{fig:Ent}
\end{minipage}\hfill
\begin{minipage}[t]{0.32\textwidth}
    \centering
    \includegraphics[width=\linewidth]{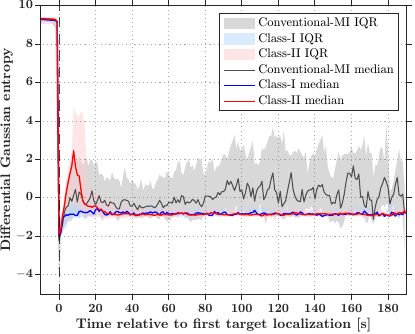}
    \caption{Posterior differential Gaussian entropy of target-state belief against the relative time for initial target localization}
    \label{fig:adjEnt}
\end{minipage}
\end{figure*}

A planar multi-agent target localization framework is adopted within a bounded search space, where four agents (1-4 from left to right) are initialized inside the domain to localize and track the target. 
The communication topology used in this paper is shown in Fig.~\ref{fig:communication network}. In the associated BAG, Agent 2 and Agent 3 are designated as AND nodes with zero external vulnerability, thereby making them attack-immune throughout the mission. Whereas Agent 1 and Agent 4 are OR nodes, also labeled as attack-prone agents, with external attack vulnerability as well as internal attack propagation through communication channels, effectively making them both prone to attack in case either is compromised.
Particles representing the target belief are generated inside the search space. The simulation parameters are shown in Table \ref{tb:margins}. Figure~\ref{fig:snaps} represents the mission evolution of multi-agent target localization. Figure~\ref{fig:s1} illustrates the initial setup of the framework, where agents are navigating inside the search space to maximize the mutual information. The initial localization of the target is seen in Fig.~\ref{fig:s2}, where Agent 4 localizes the target. It is further observed from the Fig.~\ref{fig:s3} that, after the target is localized, the false data is injected into the localizing agent, resulting in the spread of the particle belief over time. Figure~\ref{fig:s4} displays the recovery-set formation after the BAG detection, which bounds the target. Since BAG detects the FDI into Agent 4 and Agent 1, measurements from these respective agents are trust-gated during particle filter update. Post-set creation, the compromised agent goes outside the target's attacking range, and the attack-immune agents are commanded into the set to relocalize the target as seen in Fig.~\ref{fig:s5}. The framework enables mutual information maximization once it observes the particles inside the observing range of the sensing agent, and if not, the guidance law is used to guide the agent to the particle-belief mean. The relocalization of the target is observed in Fig.~\ref{fig:s6}, where both attack-immune agents relocalize the intruder as seen from the shrinkage of the particle-belief.\\
To evaluate the system's performance, a Monte Carlo study was conducted over multiple randomized target initial positions. The runs were separated into two classes: Class-I, where an attack-immune agent localizes the target initially, and no further recovery set is created or required, and Class-II, where an attack-prone agent localizes the target initially, and a recovery set is created to relocalize the target. Out of $250$ Monte Carlo runs, $146$ runs were Class-I, $104$ runs were Class-II, and none of them were unsuccessful, denoting a robust localization framework, within the tested simulation conditions. The final median target-localization times for Class-I and Class-II were $39 \, s$ and $52 \, s$ respectively. To validate the effectiveness of the proposed framework, a comparison is drawn against a conventional mutual-information-based target localization framework without BAG-based attack detection and the subsequent steps it enables.\\
For each class, the root-mean-square error (RMSE) of the target position and the posterior differential Gaussian entropy of the target belief were aggregated using the median. Figure~\ref{fig:RMSE} illustrates that the target-position RMSE for Class-I converges to approximately $0 \, m$, faster than for Class-II, which is consistent with the shorter median target-localization time. Similarly, Fig. \ref{fig:Ent} illustrates that the median posterior differential Gaussian entropy of the particle-belief converges to a constant value below the entropy threshold faster for Class-I compared to Class-II, as the recovery set is created in Class-II, which expands the entropy post-attack and re-shrinks the belief, thereby requiring more time for relocalization. Additionally, Fig.~\ref{fig:RMSE} and \ref{fig:Ent} compare proposed framework against conventional approach, under which the entropy is observed to remain relatively low despite the high RMSE, indicating that the agents chase a corrupted target belief with unwarranted confidence while the true target escapes.\\
Figure~\ref{fig:adjEnt} illustrates the posterior differential Gaussian entropy of target state belief with time relative to initial target-localization, obtained by shifting the Monte Carlo run so the initial localization instant is set to $t=0\,s$. The plot examines the uncertainty evolution of the target estimate after the first localization across two outcome classes. For Class-I the entropy decreases significantly at the instant of the first localization of the target and then converges to a constant value signifying target localization. Similarly, for Class-II, the posterior entropy drops at the time of first localization via attack-prone agents, which spikes back due to the FDI to the sole localizing agent. The entropy then decays to a constant value post-relocalization of the target via attack-immune agents during the recovery phase, signifying the relocalization of the target.

\begin{table}[t]
\begin{center}
\caption{Simulation parameters}\label{tb:margins}
\begin{tabular}{ccc}
Parameter & Symbol & Value \\ \hline
Number of particles & $ N $ & $1500$ \\
Target accel input cov & $\boldsymbol\Sigma_{\boldsymbol{\nu}_T}$  & $\mathbf{I}_2$\\
Meas. range std dev & $\sigma_r$ & $0.20 \,\mathrm{m}$\\
Meas. bearing std dev & $\sigma_\theta$ & $3^\circ$ \\
FDI attack std dev & $\sigma_\alpha$ & $2 \, \mathrm{m}$\\ 
FDI attack mean & $b$ & $15 \, \mathrm{m}$ \\
Maximum sensor range & $z_{\text{max}}$ & $10 \, \mathrm{m}$ \\
Agent max speed & $v_{\text{max}}$ & $1.5 \,\mathrm{m/s}$ \\
Target max speed & $v_{T_\text{max}}$ & $0.5 \,\mathrm{m/s}$ \\
Agent max accel & $a_{\text{max}}$ & $0.75 \,\mathrm{m/s^2}$ \\
MI grid step &  $\Delta z^r$ & $0.25 \,\mathrm{m}$\\
MI time step & $dt$ & 1 s \\   
Differential entropy threshold & $H_\mathrm{th}$ & 0 \\
LQR state error penalty & $\mathbf{Q}$  & $\text{diag}([10, 10, 5, 5])$ \\
LQR control effort penalty & $\mathbf{R}$ & $0.5\,\mathbf{I}_2$\\
\hline
\end{tabular}
\end{center}
\end{table}

\section{Conclusions}
The paper presented a resilient multi-agent target localization and tracking framework operating under limited agents' sensing range and FDI attacks. The proposed framework combined mutual information-based navigation and target search, BAG-based attack detection, and reachable set guided recovery to localize the target, and continued target tracking. The result illustrated that the proposed framework maintained localization performance even after the FDI attack on the system. Monte Carlo analysis presented two distinct localization classes: a scenario in which an attack-immune agent localized the target initially, achieving lower median localization time, compared to the scenario where an attack-prone agent localized the target initially, necessitating recovery set formation and relocalization of the target through the attack-immune agent, resulting in higher localization time, and establishing that BAG-based detection is necessary for relocalization under FDI attacks. These results illustrate the effectiveness of the proposed framework and support its relevance for the resilient target-localization applications. Future work includes incorporating sUAV dynamics for the agent to track a dynamic target and extending the framework to a decentralized system with communication constraints.

\bibliography{ifacconf}

\end{document}